**Supplemental Information for meta-analysis evaluation**

S. Stanley Young, CGStat, Raleigh NC 27607
Warren Kindzierski, School of Public Health, University of Alberta, Edmonton, Alberta T6G 1C9

This report can be found at https://arxiv.org/abs/1808.04408

**Abstract**

Massive numbers of meta-analysis studies are being published. A Google Scholar search of the phrase "systematic review and meta-analysis" returned ~452k hits since 2014. Citations were not included. The search was done Jan 14, 2019. There is a need to have some way to judge the reliability of a positive claim made in a meta-analysis that uses observational studies. Our idea is to examine the quality of the observational studies used in the meta-analysis and to examine the heterogeneity of results coming from those studies. Here we provide background information and examples: a listing of negative studies, a simulation of p-value plots, and several examples of the use of p-value plots.

**Info 00: Introduction**

This Info collection includes items useful for understanding the reliability of a meta-analysis study where the base papers use observational data. It is helpful to see a typical results presentation. Figure 1 gives a typical example diagram of the literature selection. The process starts with a computer search for possible papers to include. Usually only a small number of papers are retained. Figure 2 gives another typical example of analysis results: The papers used in the analysis with a pointer to the references along with the year of publication; In this case, the type of study; How the studies are weighted in the overall estimated effect (usually inverse variance); The risk estimates along with confidence limits; A graphical display of study results called a Forest Plot; A p-value that tests the risk level versus no effect. Along the bottom of the figure some summary statistics are given. In this case, $I^2$ is given which measures the heterogeneity of the results. The Egger regression test, which is a measure of publication bias (negative studies are much less likely to be published).

In environmental epidemiology there are many papers indicating an association between some air component and a health effect. There are a relatively small number of published negative studies, no association between an air component and a health effect. See **Info 01** for a partial list of negative papers. Note that if something is causal then there should be no valid negative studies. Multiple negative studies call causality into question.

There are multiple base studies used in a meta-analysis and each provides a risk ratio, RR, and lower and upper confidence limits, CLlow and CLhigh, that are expected to contain the RR 95% of the time. These (RR, CLlow, CLhigh) can be used to test for a difference from 1.000 (no effect) giving one p-value for each base paper. (In some cases, the no effect level is 0.00.) The ranked p-values can be plotted against the integers, 1, 2, 3,.. to give a p-value plot. If the p-value plot is roughly a 45-degree line, then the p-values are consistent with randomness, no effect. If the slope of the line is less than 45 degrees and all the points are on the line, then that is evidence for a consistent positive effect. If the pattern is bi-linear (small p-values on the blade of a hockey stick and no-effect studies on the handle of the hockey stick, _/, then that is evidence for study heterogeneity; the two results are incompatible. The small p-values indicate an effect and the larger p-values on the handle of the hockey stick indicate no effect. See **Info 02** for a simulation of p-values coming from no effect studies.

An example is helpful. In the late 1990s, there was considerable interest in $2^{nd}$ hand cigarette smoke as a cause of lung cancer. We present data from an EPA meta-analysis as an example of various plots useful for the evaluation of a meta-analysis. **See Info 03**.

**Info 04** examines air quality and cardiovascular effects.

**Info 05** gives a p-value plot for a meta-analysis that examines an association between apathy and dementia.

We are at the point where it is known that many claims made in observational papers do not replicate. This background information, Info 05, is in support of a novel point of view: Over time a false claim can become the accepted belief. Consider the following scenario:

1. Random chance or analysis manipulation gives an initial positive result, often surprising and attention gathering.
2. Editors generally accept only papers with a p-value <0.05.
3. Other researchers join in the hunt for verification of the positive result and eschew publication of negative results, Greenwald.
4. Over time, many more positive than negative papers are published.
5. A false result becomes established/canonized.

Several technical things can be used to examine the reliability of observational studies. First, the analysis search space can be counted for each paper used in a meta-analysis. Second a p-value plot can be used to examine the heterogeneity of papers used in analysis. A volcano plot plots the negative log10 of the p-value versus the measured effects and can be used to examine positive and negative effects.

There are four key references taken together support steps 1-5 above. Greenwald noted that only 6 percent of researchers were inclined to publish a negative result whereas 60 percent were inclined to publish a positive result, ~10 to 1. Head et al. (2105) noted that p-hacking (testing many questions within the same data set) is essentially universal; physics, math, and statistics appear to be exceptions. Many researchers consider testing of lots of question as business as usual, not cheating. Simonsohn et al. (2014) note that replication of a finding is not necessarily support of a claim if there is analysis or data manipulation. Both Head et al. (2105) and Simonsohn et al. (2014) use the distribution of p-values to support their claims. Finally, Nissen et al. (2016) note that it is possible for a false claim to become established truth under certain conditions that relate to the ratio of positive to negative studies.

**Info 06** points to a paper by Fraser et al. (2018) on Questionable Research Practices.

**Info 07** gives examples of p-value plots for meta-analysis studies.

**References used in these Info items:**

van Dalen JW, van Wanrooij LL, van Charante EPM, Brayne C, van Gool WA, Richard E. 2018. Association of apathy with risk of incident dementia: a systematic review and meta-analysis. JAMA Psychiatry doi:10.1001/jamapsychiatry.2018.1877

**Figure 1.** Selection of papers for inclusion.

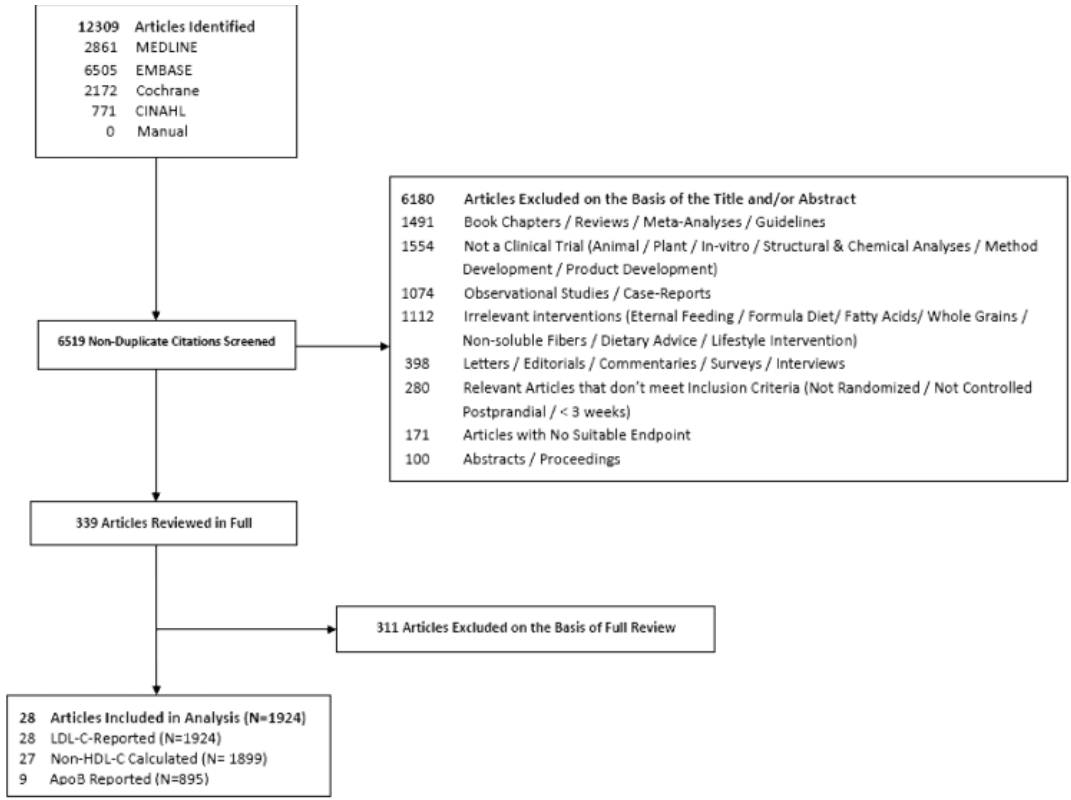

FIGURE 1 Summary of the search and selection process for the effect of psyllium on LDL-C, non-HDL-C, and apoB. Reports were identified from Medline, EMBASE, the Cochrane Central Register of Controlled Trials, CINAHL, and through manual searches of reference lists of selected studies and reviews. apoB, apolipoprotein B; LDL-C, LDL cholesterol; Non-HDL-C, non-HDL cholesterol.

**Figure 2.** Listing of studies, Weight used in computing an overall effect, Relative Risk, Confidence Intervals, a so-called Forest Plot, and a p-value for RR relative to no effect.

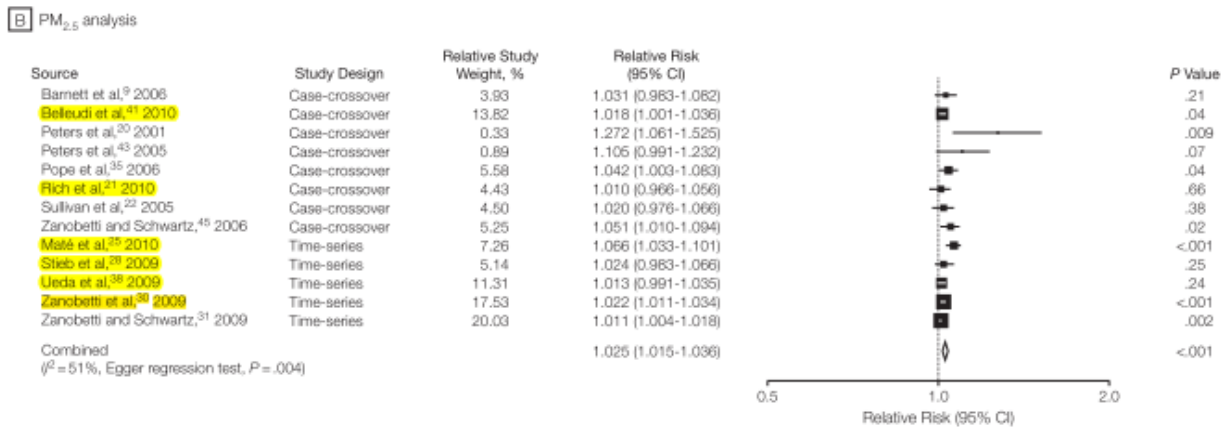

# Info 01 Negative papers

https://junkscience.com/2018/06/negative-studies-and-pm2-5/

**Negative papers are important**.

**Logic:** It takes only one <u>valid</u> negative paper to end the claim of causality coming from multiple association studies.

You can get a positive paper in several ways:
• **The reported effect is real.** People usually think this is the most likely way. In many areas of science, claimed effects only replicate 10-20% of the time.
• **The reported is spurious.** A statistical false positive due to chance. This happens rarely if there is only one question at issue. In reality, there are often hundreds of questions at issue and if that is the case, a finding/claim is much more likely to be false.
• **The reported effect is statistical fraud.** A positive due to data and/or analysis manipulation (e.g., p-hacking, trying multiple models and selecting the one you like.)

**Negative papers are a counterweight to positive papers**, but they are more difficult to get published:
• The researcher may give up and not submit a paper – the file drawer problem.
• Editors don't like negative papers — publication bias.
• Referees don't like papers against what they might have published, human prejudice.
• Researchers might think it through. If I publish a negative paper, it will be difficult to ask for follow up funding on this question!

**Negative Environmental Epidemiology References**
The studies are in chronological order with the year of publication in **bold**.

Styer, P., McMillan, N., Gao, F., Davis, J., Sacks, J., **1995**. Effect of outdoor airborne particulate matter on daily death counts. Environ. Health Perspect. 103:490-497.

Chay K, Dobkin C, Greenstone M. **2003**. The Clean Air Act of 1970 and adult mortality. Journal of Risk and Uncertainty 27:279–300.

Enstrom JE. **2005**. Fine particulate air pollution and total mortality among elderly Californians, 1973–2002. Inhalation Toxicology 17:803–816.

Janes H, Dominici F, Zeger S. **2007**. Trends in air pollution and mortality: An approach to the assessment of unmeasured confounding. Epidemiology, 2007; 18:416–423.

**Info 02 Simulated p-value plots**

A p-value plot is constructed by rank ordering the p-values from smallest to largest and plotting them against the integers, 1, 2, 3,…. If they fall on a 45-degree line, that is evidence for a uniform distribution, complete randomness. We simulate 15 p-values ten times and produce the p-value plots, Figure 1. Each of the 10 simulated meta-analysis studies has 15 base papers. Each of the base papers has only one question under consideration and any model adjustment is specified before the data is examined. The simulation is designed to represent an ideal process. The goal is to get an impression of the variability of a p-value plot under ideal conditions where there is nothing going on and there is no multiple testing or multiple modeling in the base studies.

We see the general lower left to upper right trend of the p-values. If the p-values were exactly uniform, they would be exactly on a 45-degree line. But p-values themselves have variability. The smallest p-value in a set should be exactly at the 1.00/15=0.0666 level if there were no variability. In these 10 simulations of 15 p-values we see that the smallest p-value can vary up or down from 0.0666 as chance is in play. By chance, there could be a p-value smaller than 0.05, the usual nominal statistical significance value.

Here is a listing of the smallest p-values for the 10 simulations.

|    | Study | Min(RU 0 1) |
|----|-------|-------------|
| 1  | 1     | 0.01530     |
| 2  | 10    | 0.01654     |
| 3  | 5     | 0.01736     |
| 4  | 8     | 0.04326     |
| 5  | 4     | 0.05062     |
| 6  | 7     | 0.06464     |
| 7  | 9     | 0.10827     |
| 8  | 3     | 0.12913     |
| 9  | 2     | 0.14121     |
| 10 | 6     | 0.15329     |

These numbers give a sense of how the smallest p-value in an ideal study can vary. The smallest p-value ranged from 0.0153 to 0.153 in this simulation.

Note also that although the expectation of the line is exactly a 45-degree line, again variability causes the actual data to vary off the 45-degree line.

We are interested in a hockey stick pattern, a number of small p-values followed by p-values falling roughly on a 45-degree line. With nothing going on, we would that pattern to be rather unusual. Having said that, the human eye-brain, after the fact, is easily let to claim one sort of pattern or another, size of gaps, number of gaps, blade of a hockey stick or plateau at the upper right of a figure.

The p-values, computed from data in a Lancet paper is plotted here in Figure 2, do not appear to be from a uniform distribution. Either there are real effects in some studies, or some of the studies have analysis manipulation, p-hacking. The right hand side of Figure 2 is consistent with randommess.

Figure 1. Simulation results, 10 samples of 15 p-values. Each sample of 15 represent a single meta-analysis.

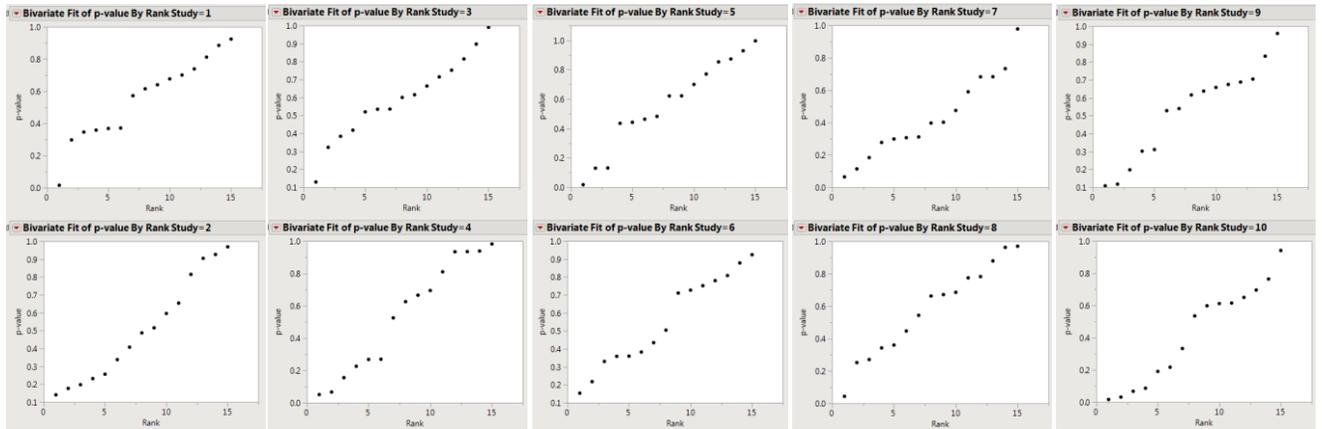

Figure 2. Plot of simulated p-values from the ten meta-analysis studies, each citing 15 base studies.

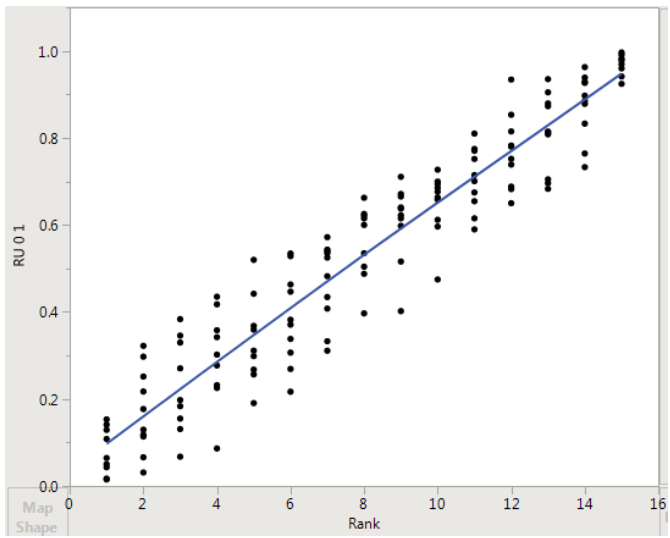

There can be a tendency to think of a p-value as a fixed, computed item. In fact, if an experiment is replicated as closely as possible a p-value will vary from one study to another. Boos and Stefanski (2013) make the point that a p-value must be quite small, 0.001, to have high assurance that a replicate experiment will have a "significant" p-value. They contend that a p-value of 0.05 is weak evidence (even under ideal conditions of no multiple testing or modeling).

Figure 3. p-values from a Lancet meta-analysis.

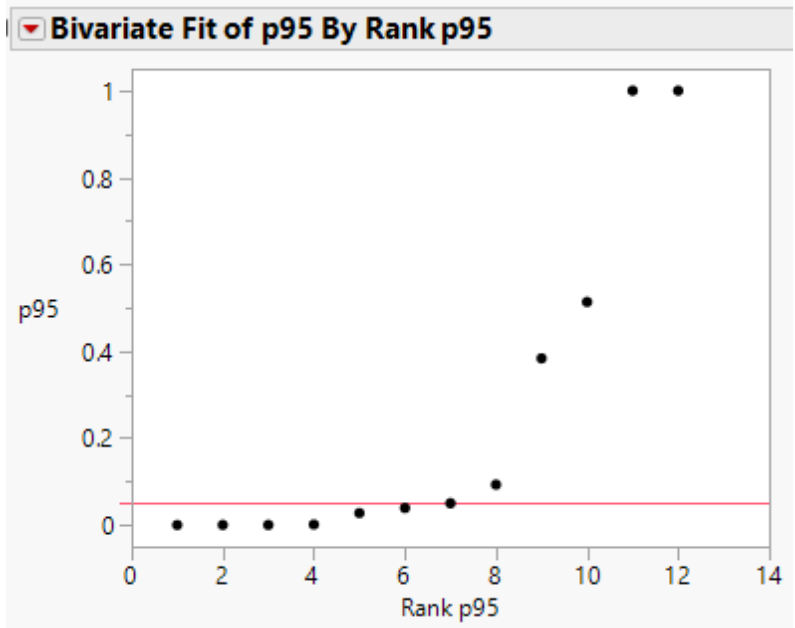

This figure gives the impression of a two-component **mixture**: small p-value studies and no effect studies, seven of each.

**Info 03 Tobacco smoke and lung cancer**

**Environmental Tobacco Smoke and lung cancer**

Gori and Luik, 1999, consider the EPA position on Environmental Tobacco Smoke (it causes lung cancer) to have been based on flawed statistical analysis, among many other problems. The EPA relied on a meta-analysis of 11 studies; see our Table 1 below which replicates the basic data in their Table 1, which appears on page 17 of the Gori and Luik book.

The EPA used 90% confidence limits rather than the usual 95% confidence limits thereby making one of the 11 CLs, Fontham, not overlap 1.000, the no effect level. None of the p-values are less than 0.05. The EPA ended up using p-value <0.10 to declare statistical significance.

P-value plots are instructive. If p-values are ranked from smallest to largest and plotted against the integers give a 45-degree line, then the data is consistent with a uniform distribution, i.e. randomness.

The negative log10 of p-values plotted against the RR is a volcano plot. The y-axis gives the credibility of a claim, larger is more unusual. The x-axis gives the magnitude of any effect, RR. The volcano plot has several features of interest. There is a gap in the middle of the figure around a RR of 1.000 that is consistent with researchers not bothering to publish a no effect study. There are more positive studies than negative, which is also consistent with publication bias. None of the p-values, expressed as -log10 p, are individually significant (0.05, blue line) or multiplicity corrected (0.00455, red line).

This case went before Federal Judge Osteen and he concluded in 1998 that the EPA had overstepped the scientific evidence.

Table 1. The variables: Author, # Cases, RR (Risk Ratio), CLlow and CLhigh (90% confidence limits, low and high) are given in Gori and Luik, page 17. From RR, CLlow and CLhigh additional statistics, standard error, z-value, p-value and the negative log10 of the p-value were computed.

| | Author | Year | Cases | RR | CLlow | CLhigh | SE90 | Z90 | p90 | Rank p90 | -log10 p |
|---|---|---|---|---|---|---|---|---|---|---|---|
| 1 | Brownson | 1987 | 19 | 1.52 | 0.49 | 4.79 | 1.307 | 0.398 | 0.691 | 10 | 0.1607 |
| 2 | Buffler | 1984 | 41 | 0.81 | 0.39 | 1.66 | 0.386 | -0.492 | 0.623 | 8 | 0.2058 |
| 3 | Butler | 1988 | 8 | 2.02 | 0.48 | 8.56 | 2.456 | 0.415 | 0.678 | 9 | 0.1688 |
| 4 | Correa | 1983 | 22 | 2.07 | 0.94 | 4.52 | 1.088 | 0.983 | 0.325 | 3 | 0.4875 |
| 5 | Fontham | 1991 | 420 | 1.29 | 1.03 | 1.62 | 0.179 | 1.617 | 0.106 | 1 | 0.9753 |
| 6 | Garfinkel1 | 1981 | 153 | 1.17 | 0.85 | 1.61 | 0.231 | 0.736 | 0.462 | 5 | 0.3356 |
| 7 | Garfinkel2 | 1985 | 134 | 1.31 | 0.93 | 1.85 | 0.280 | 1.109 | 0.268 | 2 | 0.5725 |
| 8 | Humble | 1987 | 20 | 2.20 | 0.90 | 5.50 | 1.398 | 0.858 | 0.391 | 4 | 0.4081 |
| 9 | Janerich | 1990 | 191 | 0.86 | 0.57 | 1.29 | 0.219 | -0.640 | 0.522 | 6 | 0.2820 |
| 10 | Kabat1 | 1984 | 24 | 0.79 | 0.30 | 2.04 | 0.529 | -0.397 | 0.691 | 11 | 0.1603 |
| 11 | Wu | 1985 | 29 | 1.41 | 0.63 | 3.15 | 0.766 | 0.535 | 0.592 | 7 | 0.2273 |

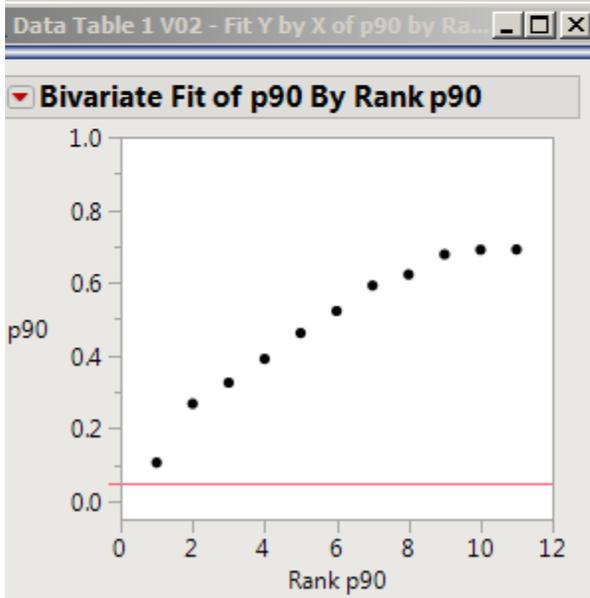

P-value plot.

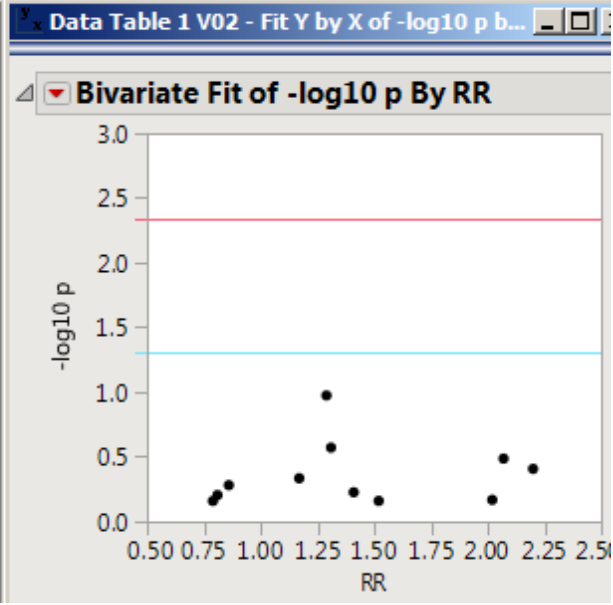

Volcano plot.

**Info 04 Air quality and cardiovascular effects**

**Introduction**

The EPA paradigm of 2016 is that air components, carbon monoxide (CO), nitrogen dioxide (NO2), particulate matter less than 10 μm in aerodynamic diameter (PM10), particulate matter less than 2.5 μm in aerodynamic diameter (PM2.5) and sulfur dioxide (SO2), and daily maximum of 8-hourly running mean of ozone, are causal of serious health effects, death, stroke, heart attacks, etc. We examine a paper by Milojevc et al. (2014) that examines these six air components and eleven (cardiovascular) health effects for hospital admissions and death. A total of 66+66=132 questions are examined. We think that multiple testing is an issue.

We present graphs to examine the Milojevc results given in their Figures 1 and 2, repeated here as our Figures 1 and 4 for hospital admissions and death respectively.

**Data**

Over 400,000 myocardial infarction (MI) events from the Myocardial Ischaemia National Audit Project (MINAP) database, over 2 million CVD emergency hospital admissions and over 600,000 CVD deaths.

The air components were measured at the nearest air pollution monitoring site to the place of residence. Pollutant effects were modelled using lags up to 4 days and adjusted for ambient temperature and day of week.

The risk ratios and confidence limits shown in Figures 1 and 4 were digitized and placed in data tables. Additional statistics were computed from these three numbers, Standard Error, Z-value, p-value. These tables are available on request.

**Methods**

When many questions are examined, each question can provide a p-value. The distribution of these p-values can be examined using a p-value plot, Schweder and Spjøtvoll (1982). If there is no effect p-values should follow a uniform distribution, fall anywhere in the interval 0 to 1. In a p-value plot, the p-values are rank ordered from smallest to largest and plotted against the integers, 1, 2, 3, …, n. If the p-values form a 45-degree line, then that is evidence for randomness, no effect. The shape of the p-value plot can indicate analysis manipulation, Simonsohn et al. (2014).

It is useful to examine p-values in conjunction with effect sizes, a volcano plot, Cui 2003. Here, the negative of the log base 10 of a p-value is plotted against the calculated effect size. P-values that spew high left and right have small p-values and large effects. The plot facilitates seeing important effects in the context of all the comparisons at issue.

**Results**

Figures 1 and 2, repeated from the Milojevic paper, give an impression of randomness with most error bars overlapping the no effect value of 1.000.

Figure 1. Hospital admissions, 6 air components and 11 health effects.

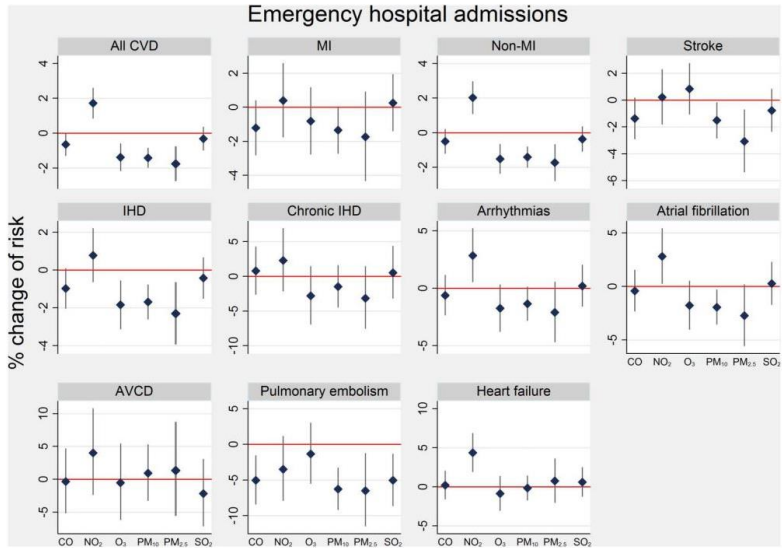

Figure 2. Deaths, 6 air components and 11 health effects.

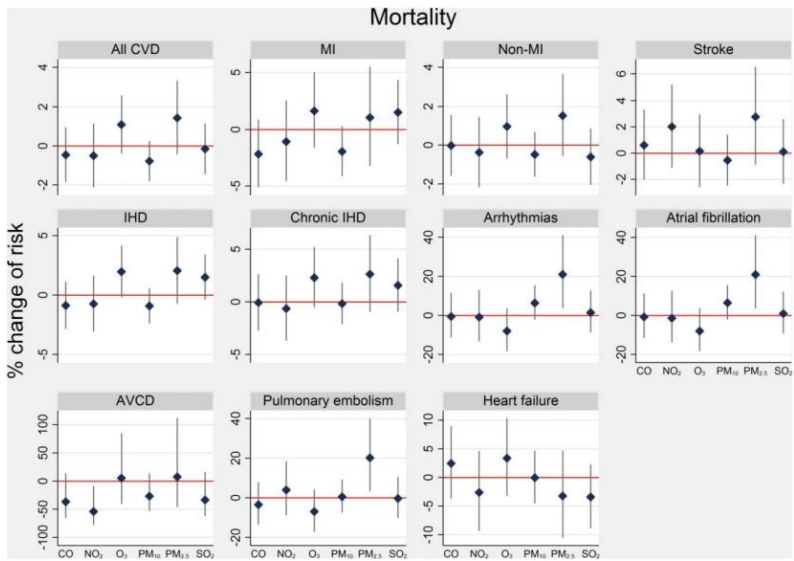

Figure 3 gives a p-value plot for 66 hospital admissions tests. There are many small p-values, which imply that hospital admissions are affected by air components.

Figure 3. p-value plot for 66 hospital admissions tests.

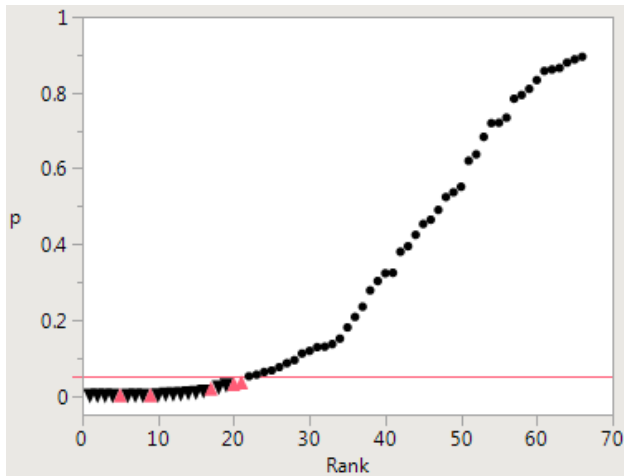

Most of the p-values <0.05 indicate a decrease in admissions. The five p-values marked in red are for increases. The figure has the appearance of p-hacking, but the authors report all effects examined and do not point out the many significant negative effects.

Figure 4, a volcano plot, shows that most of the significant effects are beneficial (a decrease in admissions). This result is surprising and we have no explanation.

Figure 4. Volcano plot for 66 hospital admissions tests.

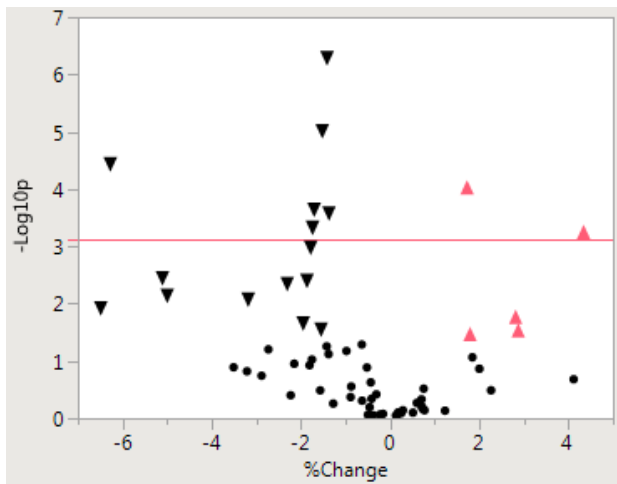

Reference line at negative log10 (0.05/66) = 3.12, the Bonferroni adjusted p-value. One outlier removed for graphical clarity. In a volcano plot, one expects the small p-values to be associated with large positive and negative effects. It seems unusual to see the spike of decreases around negative 2.0.

Figure 5 P-value plot for 66 deaths.

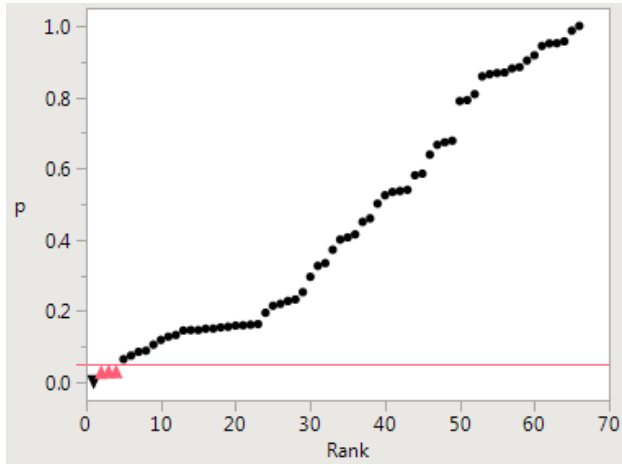

There are four p-values <0.05, three for increases.

Figure 6. Volcano plot for deaths.

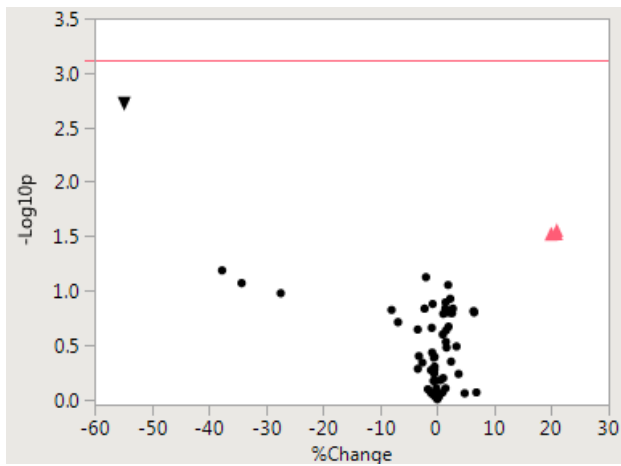

Reference line at negative log10 (0.05/66) = 3.12.

Again, it is unexpected to see the spike in small p-values in and around the no effect level of zero. For Figures 5 and 6, deaths, there are many more p-values indicating beneficial effects than harmful effects.

**Conclusion**

The authors state, "This study found no clear evidence for pollution effects on STEMIs and stroke, which ultimately represent thrombogenic processes, though it did for pulmonary embolism." Given the number of tests at issue, the claim of pulmonary embolism could well be chance. We conclude that the Milojevic study is essentially negative.

**Info 05 Apathy and dementia**

**P-value plot for van Dalen meta-analysis**

Van Dalen et al. (2018) present a meta-analysis of studies that examine an association of apathy with dementia. We compute a p-value for each of the 12 studies used, rank the p-values from smallest to largest and plot them against the integers, 1, 2, …,12. Data taken from van Dalen is given in Table 1. P-value calculations are also given in Table 1.

If there are no effects, the p-value plot will be linear, 45-degree line. If there is a consistent effect, the plot will be linear with a slope less than 45 degrees. If the results are heterogeneous, then there will be a bilinear response, p-values on the left will have a shallow slope and p-values on the right will have a 45-degree line.

We observe what appears to be a bilinear effect.

The p-value plot is given as Figure 1. A quadratic fit is substantially better than a simple linear fit, p<0.004, Figure 1.

Comment: The gap in Figure 1 between points 8 and 9 could be chance, but it is also consistent with negative studies not being published.

Table 1. Data taken from van Dalen, Figure 1.

| RowID | Author | Ref | Year | RR | CLlow | CLhigh | SE95 | Z95 | p95 | Rank p95 |
|---|---|---|---|---|---|---|---|---|---|---|
| 1 | 1 Pink | 35 | 2015 | 1.37 | 0.98 | 1.91 | 0.2372 | 1.5596 | 0.1189 | 5 |
| 2 | 2 Robert | 28 | 2008 | 1.57 | 1.00 | 2.46 | 0.3724 | 1.5304 | 0.1259 | 6 |
| 3 | 3 Somme | 33 | 2013 | 1.13 | 0.66 | 1.94 | 0.3265 | 0.3981 | 0.6905 | 9 |
| 4 | 4 Sobow | 34 | 2014 | 6.30 | 3.38 | 11.74 | 2.1327 | 2.4852 | 0.0129 | 2 |
| 5 | 5 Peters | 23 | 2013 | 0.91 | 0.48 | 1.71 | 0.3138 | -0.2868 | 0.7742 | 10 |
| 6 | 6 Chilovi | 32 | 2009 | 2.12 | 1.12 | 3.99 | 0.7321 | 1.5298 | 0.1261 | 7 |
| 7 | 7 Palmer | 36 | 2010 | 4.83 | 2.09 | 11.18 | 2.3189 | 1.6517 | 0.0986 | 4 |
| 8 | 8 Teng | 31 | 2007 | 2.92 | 1.14 | 7.48 | 1.6173 | 1.1871 | 0.2352 | 8 |
| 9 | 10 Chan | 39 | 2011 | 0.38 | 0.12 | 1.18 | 0.2704 | -2.2928 | 0.0219 | 3 |
| 10 | 11 Brodaty | 37 | 2012 | 1.47 | 0.09 | 23.41 | 5.9490 | 0.0790 | 0.9370 | 11 |
| 11 | 12 Burke | 30 | 2016 | 2.00 | 1.78 | 2.25 | 0.1199 | 8.3404 | 0.0000 | 1 |

Figure 1. P-value plot of van Dalen data set.

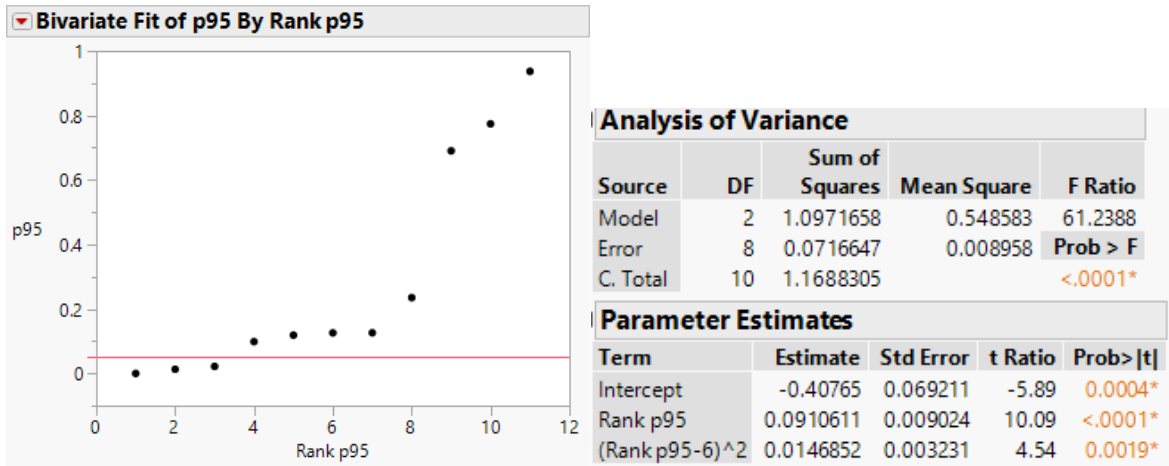

**Info 06: Questionable Research Practices**

The paper by Fraser et al. (2018) does not paint a pretty picture of ecology and evolution.

Fraser et al. (2018) is worth a read. Here is their Abstract (italics added):

> We surveyed 807 researchers (494 ecologists and 313 evolutionary biologists) about their use of Questionable Research Practices (QRPs), including *cherry picking statistically significant results, p hacking, and hypothesising after the results are known (HARKing).* We also asked them to estimate the proportion of their colleagues that use each of these QRPs. Several of the QRPs were prevalent within the ecology and evolution research community. Across the two groups, we found *64% of surveyed researchers reported they had at least once failed to report results because they were not statistically significant (cherry picking);* 42% had collected more data after inspecting whether results were statistically significant (a form of p hacking) and *51% had reported an unexpected finding as though it had been hypothesised from the start (HARKing).* Such practices have been directly implicated in the *low rates of reproducible results* uncovered by recent large scale replication studies in psychology and other disciplines. The *rates of QRPs found in this study are comparable with the rates seen in psychology*, indicating that the reproducibility problems discovered in psychology are also likely to be present in ecology and evolution.

They describe science as transforming "ugly initial results into beautiful articles". Everyone likes a good story.

There are repeated failures to reproduce a large proportion of results in the published literature. What are the ways in which this transformation, ugly to beautiful, **process is susceptible to confusion and <u>corruption</u>?**

Researchers have a very successful business model:
1. Ask a lot of questions (p-hacking) (also sculpt data and change statistical models)
2. Make up a rationalization after the fact (HARKing); tell a good story.
3. Do not give up their data set to anyone.
Many researchers have no "skin in the game". They are using other people's money and their future livelihood does not depend to the validity of claims made. Are they trustworthy?

**07 Info** Examples of meta-analysis p-value plots.

1. Lancet PM10/PM2.5.

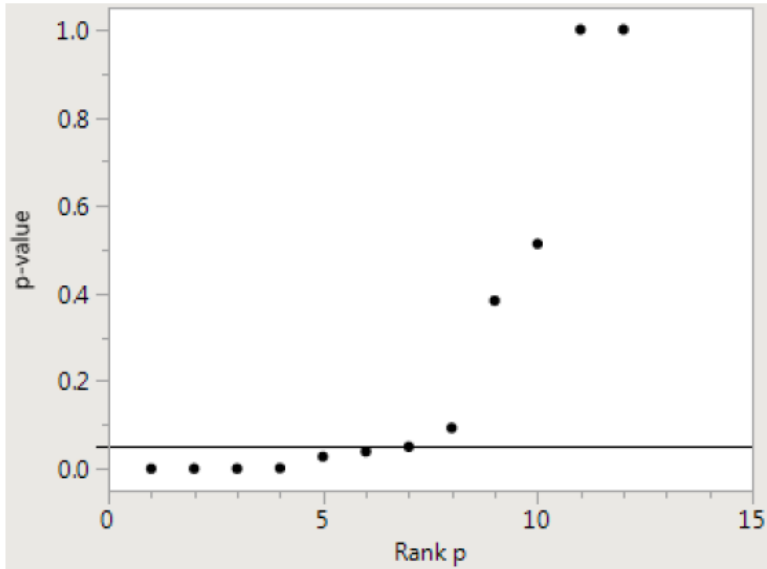

2. JAMA paper.

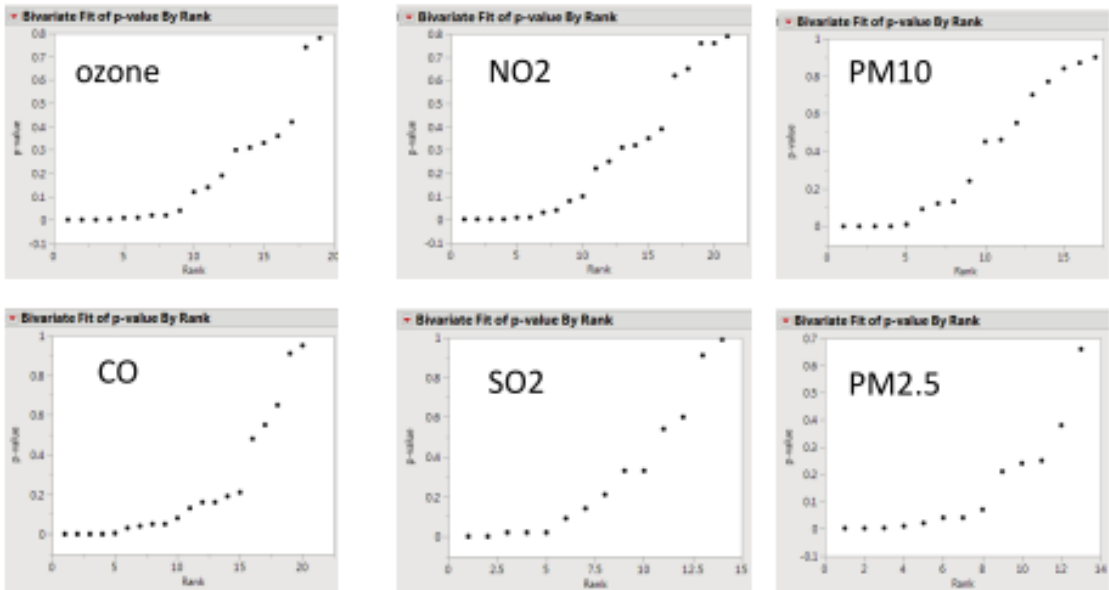

For each component there are many not significant results.

3. JAMA Open meta-analysis of RCTs.

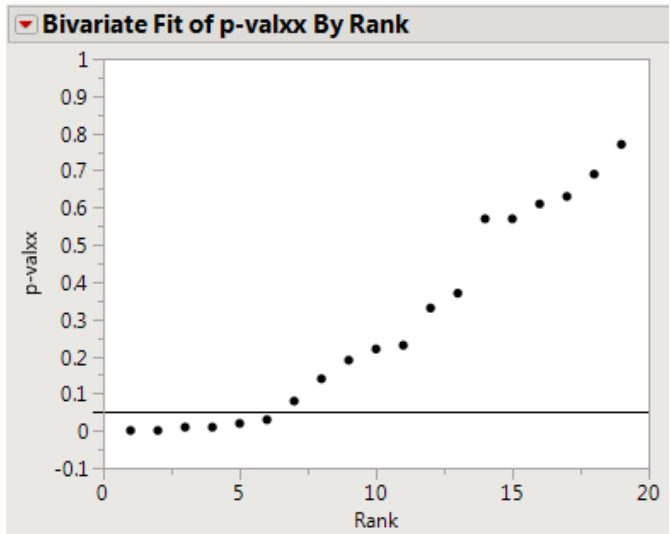

4. Nutrition study 1, American Journal of Clinical Nutrition.

Figure 1                                                                 Figure 2

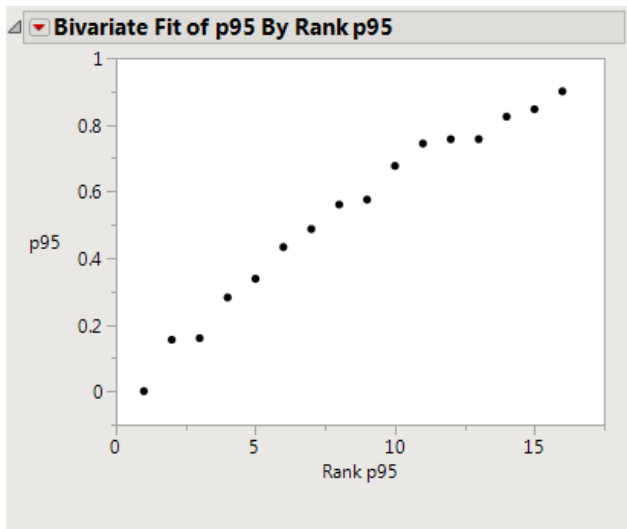 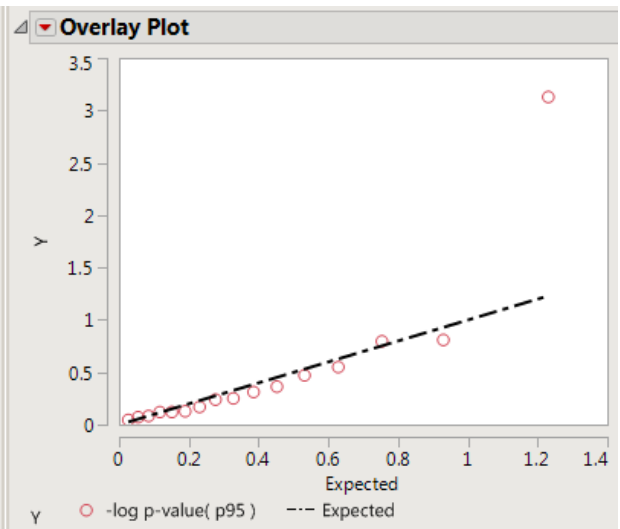

The authors claim an overall effect, which appears questionable/unreasonable from Figure 1. All but one of the p-values are non-significant. The authors fail to note one extremely small p-value, 0.00074348, that completely dominates the overall effect. Figure 2 plots the -log10 of the p-value against its expected value. All the points but one fall on the no effect line. The study with the very small p-value merits examination.

5. Nutrition Study 2, food frequency questionnaire study.

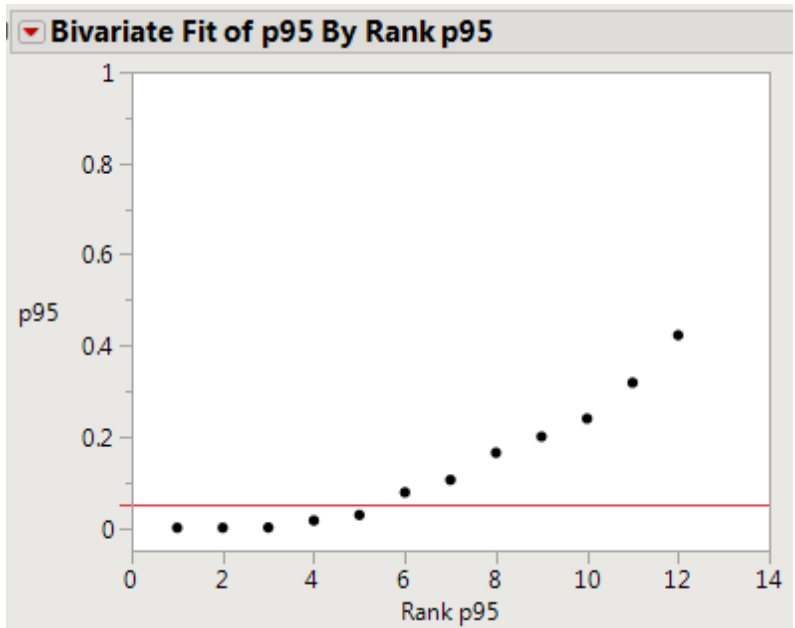

The typical food frequency questionnaire study will have 60-200 foods at issue. Common practice is to write a paper around a nominally statistically significant result. Here we see the typical hockey stick p-value plot but <u>with no p-values above ~0.40</u>. What is going on??? There could well be reporting bias in that many studies looked at the food at issue in this study but reported on other foods and did not report on the nonsignificant food at issue on this study. The smallest two p-values reported were 0.00000859784 and 0.000057423. If those p-values are real, then it is very unlikely to get a p-value as large as 0.42265814. We appear to have a mixture.

6. Example from PLOS 2015, polychlorinated biphenyl (PCB).

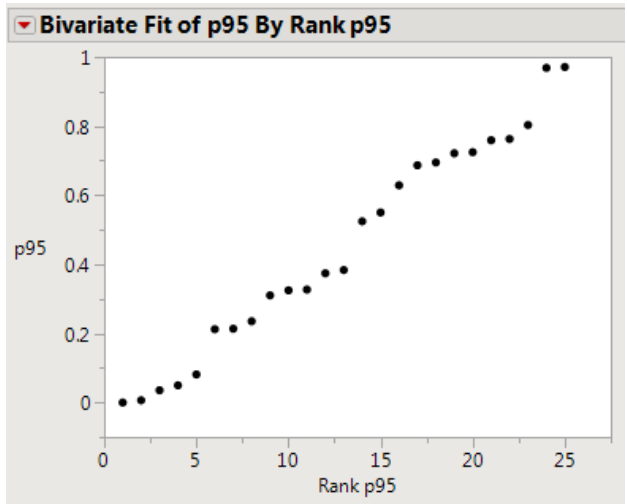

The general pattern is of points falling on a 45-degree line indicating no effect. There are two very small p-values, 0.00004065 and 0.0066507, which would normally be taken as indicating a real effect but for the many non-significant findings. The authors look at various subgroups of the studies and conclude "Our meta-analysis based on the *selected* studies found group II and group III PCB exposure *might* contribute to the risk of breast cancer." Italics added.